\documentclass[conference]{IEEEtran}
\IEEEoverridecommandlockouts

\usepackage{marvosym}
\usepackage[numbers,sort&compress]{natbib}
\usepackage{hyperref}
\hypersetup{
    colorlinks=true,    
    linkcolor=blue,     
    citecolor=blue,     
    filecolor=magenta,  
    urlcolor=cyan,      
}
\newcommand{\figref}[1]{\hyperref[#1]{Fig.~\ref*{#1}}}
\newcommand{\secref}[1]{\hyperref[#1]{section~\ref*{#1}}}
\newcommand{\tabref}[1]{\hyperref[#1]{Table~\ref*{#1}}}

\usepackage{amsmath,amssymb,amsfonts}
\usepackage{algorithmic}
\usepackage{graphicx}
\usepackage{textcomp}
\usepackage{xcolor}
\usepackage{tabularx} 
\def\BibTeX{{\rm B\kern-.05em{\sc i\kern-.025em b}\kern-.08em
    T\kern-.1667em\lower.7ex\hbox{E}\kern-.125emX}}
\begin{document}

\title{STFTCodec: High-Fidelity Audio Compression through Time-Frequency Domain Representation}
% \author{Anonymous ICME submission}
\author{
    \IEEEauthorblockN{
        Tao Feng$^{1,2}$\IEEEauthorrefmark{1}, 
        Zhiyuan Zhao$^2$\IEEEauthorrefmark{1}, 
        Yifan Xie$^1$, 
        Yuqi Ye$^3$, 
        Xiangyang Luo$^{1,2}$, 
        Xun Guan$^1$\IEEEauthorrefmark{2},
        Yu Li$^2$\IEEEauthorrefmark{2}
    }
    \vspace{0.1cm}
    
    \IEEEauthorblockA{
        $^1$Shenzhen International Graduate School, Tsinghua University, Shenzhen, China
    }
    \IEEEauthorblockA{
        $^2$International Digital Economy Academy (IDEA), Shenzhen, China
    }
    \IEEEauthorblockA{
        $^3$Shenzhen Graduate School, Peking University, Shenzhen, China\\
    \vspace{0.1cm}
    Email: ft23@mails.tsinghua.edu.cn}
    \thanks{This work was done during Tao's Intern at IDEA.}
    \thanks{$^*$Equal contribution. $^{\dag}$Corresponding authors.}
}

\maketitle

\begin{abstract}
    We present STFTCodec, a novel spectral-based neural audio codec that efficiently compresses audio using Short-Time Fourier Transform (STFT). Unlike waveform-based approaches that require large model capacity and substantial memory consumption, this method leverages STFT for compact spectral representation and introduces unwrapped phase derivatives as auxiliary features. Our architecture employs parallel magnitude and phase processing branches enhanced by advanced feature extraction mechanisms. By relaxing strict phase reconstruction constraints while maintaining phase-aware processing, we achieve superior perceptual quality. Experimental results demonstrate that STFTCodec outperforms both waveform-based and spectral-based approaches across multiple bitrates, while offering unique flexibility in compression ratio adjustment through STFT parameter modification without architectural changes.
\end{abstract}


\section{Introduction}
\label{sec:intro}
Neural audio codecs leverage deep neural networks for audio compression, serving the fundamental purpose of compressing audio signals into compact discrete representations while maintaining high perceptual quality \cite{wu2024codec}. Their importance is further emphasized as these models can effectively encode continuous audio signals into discrete tokens through vector quantization, enabling their role as tokenizers in audio language models \cite{borsos2023audiolm, wang2023neural} and various audio generation tasks \cite{ju2024naturalspeech, agostinelli2023musiclm}.

The field has evolved from traditional parametric codecs (e.g., Opus \cite{valin2012definition} and EVS \cite{dietz2015overview}) to neural-based approaches. A significant breakthrough came with VQ-VAE \cite{van2017neural} and its application to speech coding \cite{garbacea2019low}. SoundStream \cite{zeghidour2021soundstream} emerged as a pioneering end-to-end waveform-based codec, establishing a paradigm combining encoder-decoder architectures with vector quantization. Building upon this architecture, EnCodec \cite{defossez2022high} introduces multi-scale STFT discriminator and spectral reconstruction losses to mitigate artifacts while accelerating training. DAC \cite{kumar2024high} further advanced this framework by incorporating factorized and L2-normalized codes to address codebook collapse, while AudioDec \cite{wu2023audiodec} enhanced it through a two-stage optimization strategy integrated with an advanced vocoder.

\begin{figure}[t]  
    \centering
    \includegraphics[width=\columnwidth]{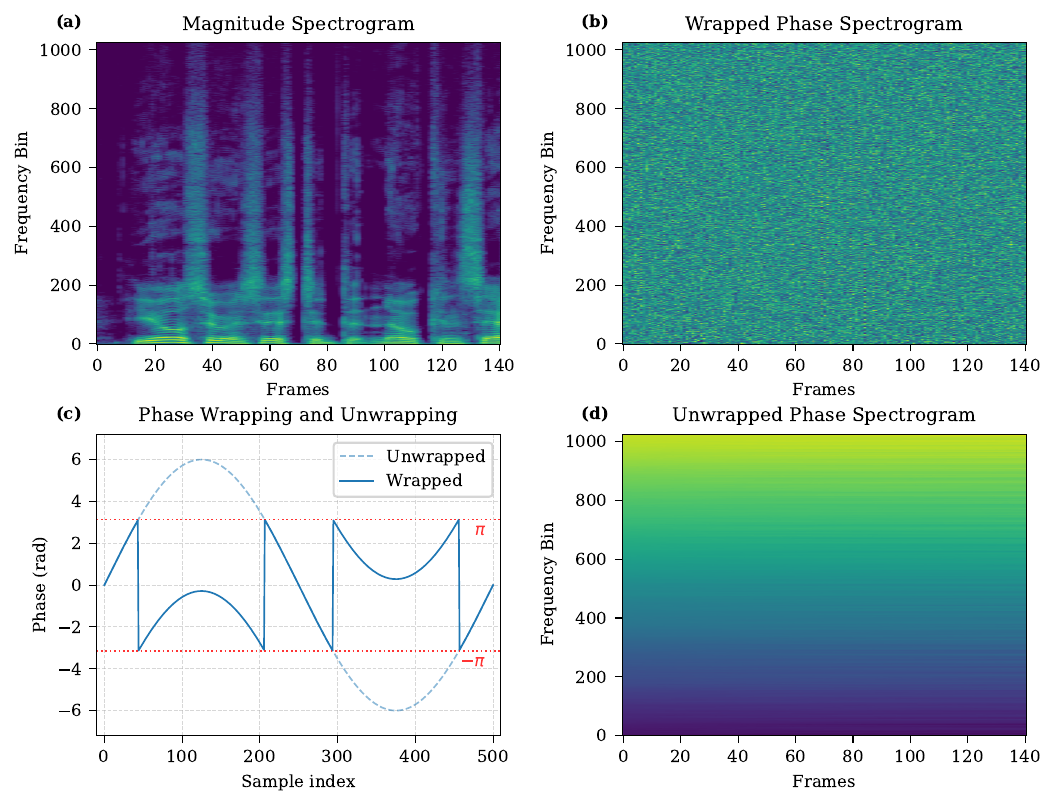}
    \caption{STFT components and phase unwrapping visualization. (a) Magnitude spectrogram. (b) Original wrapped phase spectrogram in [-$\pi$, $\pi$]. (c) Phase wrapping illustration with wrapped phase (blue solid) and unwrapped phase (blue dashed). (d) Fully unwrapped phase spectrogram after frequency-wise and time-wise unwrapping.}
\label{fig:stft}
% \vspace{-1.3em}
\end{figure}

However, as pointed out by \cite{siuzdak2023vocos}, direct modeling of raw audio waveforms faces two critical challenges: (1) Time-domain signals contain significant redundancy and high-frequency details, requiring models with substantial capacity for effective representation, and (2) the computational complexity grows linearly with the sequence length, leading to increased memory consumption and computational overhead, particularly for high-sampling-rate audio (e.g., 44.1kHz or 48kHz).

While recent works \cite{siuzdak2023vocos, ai2024apcodec, jiang2024mdctcodec} have explored spectral-based approaches to address these limitations, our method uniquely prioritizes perceptual quality over strict spectral reconstruction, featuring an enhanced architecture with advanced feature processing. Our contributions are summarized as follows:
% \vspace{-1em}
\begin{itemize}
    \item We propose a novel neural audio codec that leverages STFT for efficient audio compression, incorporating unwrapped phase derivatives as auxiliary features.
    \item We design an efficient neural architecture with separate magnitude and phase processing branches, enhanced by advanced feature extraction and attention mechanisms.
    \item We demonstrate that relaxing strict phase reconstruction constraints leads to better perceptual quality while maintaining phase-aware audio reconstruction.
    \item We show the unique flexibility of our STFT-based approach in achieving higher compression ratios through simple hop size adjustment without architectural changes.
\end{itemize}

\section{Related work}

\subsection{Time-Frequency Transforms as Downsampling Layers}
Short-Time Fourier Transform (STFT), a fundamental time-frequency representation in audio processing, serves as an efficient alternative to conventional downsampling layers. Unlike traditional downsampling methods requiring cascaded operations, STFT naturally compresses temporal information through its windowed analysis while guaranteeing perfect reconstruction. Vocos \cite{siuzdak2023vocos} demonstrated this advantage by replacing multiple upsampling layers with one inverse STFT operation, achieving competitive audio quality with substantially improved efficiency.

A key challenge in utilizing STFT is phase modeling, due to its periodic structures and wrapping behavior within [-$\pi$, $\pi$] interval (as shown in \figref{fig:stft}(c)). Various approaches have been proposed to address this challenge. Vocos \cite{siuzdak2023vocos} employed circular activation functions to generate phase information instead of direct reconstruction, effectively addressing the difficulties in phase modeling caused by phase wrapping. Building upon this, APCodec \cite{ai2024apcodec} carefully designed physically-informed phase losses (including anti-wrapping instantaneous phase loss, anti-wrapping group delay loss, and anti-wrapping instantaneous angular frequency loss) to further constrain the generated phase components to follow natural phase patterns. These losses specifically address the periodic nature and continuity properties of phase in time-frequency domain. Meanwhile, TiFGAN \cite{marafioti2019adversarial} took a different approach by suggesting unwrapped phase derivatives as model targets rather than the phase itself, which helps avoid the wrapping issue.

\subsection{Spectral-based Neural Audio Codecs}
Time-frequency transforms have long been the cornerstone of audio processing. Recently, \cite{davidson2023high} and \cite{lim2023end} demonstrated that combining these traditional methods with neural networks, particularly through MDCT, can achieve competitive or superior performance compared to traditional codecs, especially at low bitrates. Building upon this success, APCodec \cite{ai2024apcodec} enhanced STFT-based coding through physically-informed spectral losses, while MDCTCodec \cite{jiang2024mdctcodec} leveraged MDCT to avoid phase-related challenges inherent in STFT methods, maintaining spectral loss to ensure high-quality MDCT spectra reconstruction.

Our approach diverges from these methods in a fundamental way. While previous works \cite{ai2024apcodec, jiang2024mdctcodec} emphasized accurate spectral reconstruction through carefully designed spectral losses, we prioritize perceptual audio quality through multi-scale mel-spectrograms. By removing all intermediate spectral constraints, our model directly optimizes for human-perceived audio quality rather than spectral accuracy.

\begin{figure}[t]  
    \centering
    \includegraphics[width=\columnwidth]{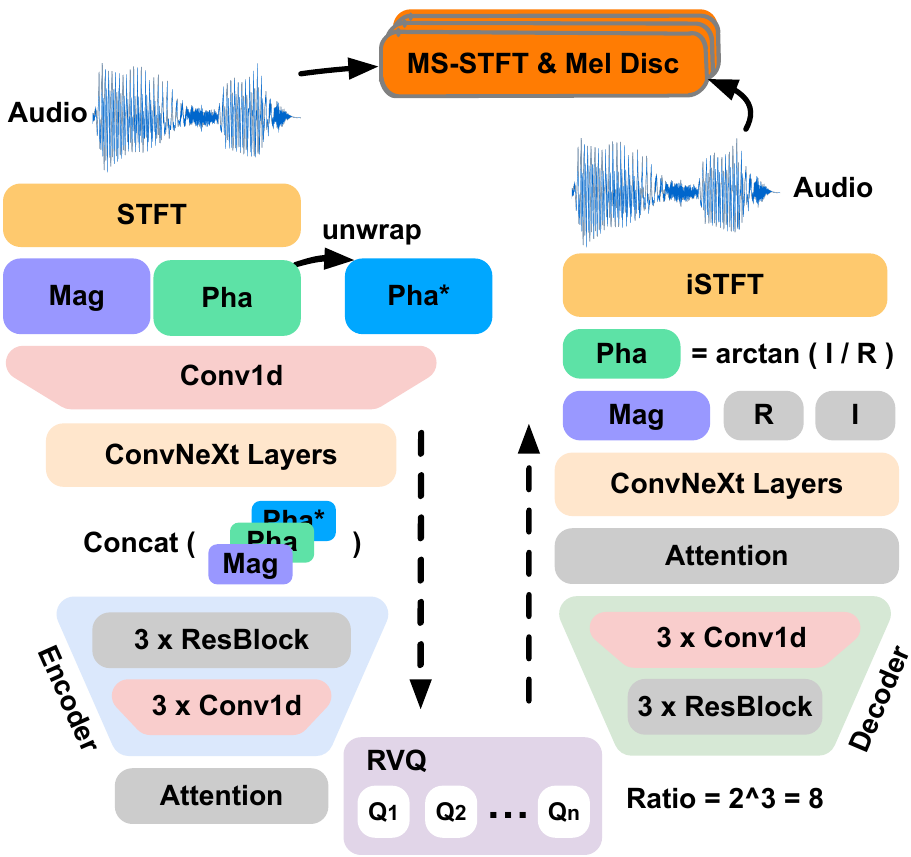}
    \caption{Model architecture with dual-branch encoder processing magnitude and phase features. The encoder compresses input through ConvNeXt layers and ResBlocks, with an total 8$\times$ downsampling ratio. After decoder upsampling, the features are processed through ConvNeXt layers, followed by convolutional layers to reconstruct magnitude spectrum and phase components for inverse STFT synthesis.}
\label{fig:model_struc}
% \vspace{-1.3em}
\end{figure}

\section{Methodology}
\label{sec:method}

We adopt the prevalent GAN framework leveraging a carefully designed encoder-decoder structure with vector quantization and multi-perspective discriminators. 

\subsection{Spectral Representation and Phase Unwrapping}
\label{sec:stft_and_unwrapping}
Unlike approaches that operate directly on time-domain waveforms, we derive our input features from the time-frequency domain through STFT. The STFT of a time-domain signal \( x[n] \) is computed as:

\begin{equation}
    STFT_{x}(m,k) = \sum_{n=0}^{N-1} x[n]w[n-m]e^{-j2\pi\frac{kn}{N}}
\label{eq:stft}
\end{equation}

\noindent where \(m\) and \(k\) are time frame and frequency bin indices respectively, \(w[n]\) is an analysis window, and \(N\) is the FFT size. 

From these complex STFT coefficients, we extract magnitude and phase spectrograms as shown in \figref{fig:stft}(a) and (b). The magnitude spectrogram represents the energy distribution across time and frequency, while the phase spectrogram exhibits discontinuities by wrapping between -$\pi$ and $\pi$ (\figref{fig:stft}(c)).

To obtain more stable features, we first unwrap the phase along the time dimension. For each frequency bin \(k\), the unwrapped phase \(\phi'[m,k]\) is computed by correcting the 2$\pi$ jumps in consecutive time frames:

\begin{equation}
% \vspace{-0.1em}
    \phi'[m,k] = \phi[m,k] + 2\pi n[m,k]
\label{eq:time_unwrap}
\end{equation}

\noindent where \(n[m,k]\) is the number of 2$\pi$ jumps needed to correct the discontinuity at time frame \(m\). We then compute the temporal gradient of the unwrapped phase:

\begin{equation}
    \nabla\phi[m,k] = \begin{cases}
        0, & m = 0 \\
        \phi'[m,k] - \phi'[m-1,k], & m > 0
    \end{cases}
\label{eq:unwrap_grad}
\end{equation}

While \figref{fig:stft}(d) demonstrates phase unwrapping along both frequency and time dimensions, we find time-wise unwrapping alone provides sufficient phase information. The resulting temporal gradient serves as auxiliary input to our model, helping to maintain phase coherence during reconstruction.

\subsection{VQ-VAE Generator}
\label{sec:vqvae_g}

\subsubsection{Encoder}
\label{sec:encoder}
Our encoder transforms the spectral representations into compact features through a hierarchical architecture. As shown in \figref{fig:model_struc}, the encoder processes magnitude spectrogram and phase information in parallel streams, with the temporal gradient of unwrapped phase serving as auxiliary input to enhance phase coherence.

Given an input signal with FFT size of 1024 and hop size of 40, the STFT analysis yields magnitude and phase spectrograms with 513 frequency bins along the temporal axis. These spectral features are first mapped to learned embeddings through convolutional layers. We set the internal dimension to 256 channels for the magnitude stream, while both the phase and phase gradient streams utilize 128 channels. These embeddings are then processed through ConvNeXt V2 \cite{woo2023convnext} blocks featuring depth-wise separable convolutions and global response normalization.

The processed features from all three streams are concatenated along the feature dimension, producing a 512-channel representation. This combined feature is fed into a unified downsampling network consisting of residual blocks and self-attention layers. Through strided convolutions, the network reduces temporal resolution by a factor of 8 while maintaining the feature dimensionality, outputting compressed features with preserved spectral characteristics.

\subsubsection{Vector quantization}
\label{sec:vq}
The encoded features are first projected to 8-dimensional space and then discretized using a residual vector quantizer (RVQ) with multiple sequential codebooks. Each codebook quantizes the residual error from previous stages. After quantization, the features are projected back to the original 512-dimensional space for subsequent decoding.

To address common challenges in vector quantization, we adopt several strategies: (1) factorized codes and L2-normalized codes \cite{kumar2024high} to achieve 100\% codebook utilization without requiring EMA-based codebook restart for dead entries \cite{dhariwal2020jukebox}, (2) variable-sized codebooks while maintaining the same total bit occupancy as a uniform 1024-size setting, enabling better utilization of the codebook space, and (3) straight-through estimator for gradient backpropagation during training, as the argmin operation in vector quantization is non-differentiable.

\subsubsection{Decoder}
The decoder mirrors the encoder's architecture to reconstruct high-fidelity audio from quantized representations. Unlike Vocos \cite{siuzdak2023vocos} which directly splits features along channel dimension or APCodec \cite{ai2024apcodec} which employs dual-stream upsampling, our decoder first employs a unified upsampling network to restore the temporal resolution from \(T/8\) to \(T\) while maintaining the 512-dimensional feature space. This unified network consists of ResNet \cite{resnet} blocks interleaved with transposed convolutions and self-attention mechanisms.

After upsampling, we utilize two parallel convolutional networks to extract magnitude and phase information from the unified features. Each extracted feature is then processed through a series of ConvNeXt v2 blocks with LayerNorm and internal dimension of 256.

For magnitude reconstruction, the network directly projects the processed features to the original frequency dimension 513 to predict log-magnitude spectrograms. Inspired by \cite{siuzdak2023vocos}, our phase reconstruction adopts an indirect approach: the phase features first undergo a linear projection followed by two parallel convolutions to predict real (\(R\)) and imaginary (\(I\)) components, from which we compute the phase angle:

\begin{equation}
\hat{\phi} = \operatorname{arctan2}(I, R)
\end{equation}

The time-domain audio signal is then recovered through inverse STFT:

\begin{equation}
\hat{x} = \text{iSTFT}(A \cdot \exp(j\hat{\phi}))
\end{equation}

\noindent where \(A = \exp(\hat{A})\) represents the magnitude spectrum obtained from the predicted log-magnitude \(\hat{A}\).

This architecture design ensures effective feature extraction while maintaining spectral coherence, and the indirect phase reconstruction strategy effectively avoids the challenges of direct phase prediction.

\subsection{Multi-Perspective Discriminators}
Rather than introducing novel discriminators, we rely on previously validated approaches that have been widely applied and shown to be effective in audio generation tasks. Specifically, we incorporate two established discriminator types: the multi-period discriminator (MPD), which was introduced in HiFi-GAN \cite{kong2020hifi} and focuses on capturing periodic patterns from raw waveforms, and a multi-scale STFT-based discriminator (MS-STFT) as used in EnCodec \cite{defossez2022high}, which evaluates spectral structures across multiple resolutions. Previous studies, including DAC \cite{kumar2024high} and BigCodec \cite{xin2024bigcodec}, report that the combination of MPD and MS-STFT discriminators yields strong perceptual quality without significant gains from adding a multi-scale waveform discriminator (MSD) originally proposed in MelGAN \cite{kumar2019melgan}. These two proven discriminators provide stable and reliable adversarial guidance for our model.

\begin{table*}[!t]
\centering
\caption{Comparison of Different Neural Audio Codec Architectures across Multiple Bitrates. The \textbf{bold} and
\underline{underline} numbers indicate optimal and sub-optimal results, respectively.}
\label{table:codec_comparison}
\begin{tabular}{lcccccccc}
\hline
\textbf{Model} & \textbf{Type} & \textbf{Bitrate} & \textbf{UTMOS $\uparrow$} & \textbf{PESQ $\uparrow$} & \textbf{STOI $\uparrow$} & \textbf{V/UV F1 $\uparrow$} & \textbf{ViSQOL $\uparrow$} & \textbf{LSD $\downarrow$} \\
\hline
GT          & - & - & 4.0305 & - & - & - & - \\
AudioDec  \cite{wu2023audiodec} & Waveform & 12.8 kbps & 3.7572 & 2.5575 & 0.8393 & 0.9302 & 4.1227 & 0.8234 \\
DAC       \cite{kumar2024high}  & Waveform & 12 kbps   & \textbf{4.0115} & \underline{4.1464} & \underline{0.9565} & \underline{0.9804} & 4.1309 & 0.7996 \\
APCodec   \cite{ai2024apcodec}  & Spectral & 12 kbps   & 3.8786 & 3.2155 & 0.9024 & 0.9642 & 4.2240 & \underline{0.7911} \\
MDCTCodec \cite{jiang2024mdctcodec}  & Spectral & 12 kbps   & -      & -      & 0.9340 & -      & \underline{4.3200} & 0.8120 \\
STFTCodec (ours)   & Spectral & 12 kbps   & \underline{3.9750} & \textbf{4.2753} & \textbf{0.9756} & \textbf{0.9865} & \textbf{4.3518} & \textbf{0.7717} \\

\hline
DAC       \cite{kumar2024high}       & Waveform & 9 kbps    & \textbf{3.9770} & \underline{3.9223} & \underline{0.9432} & \underline{0.9771} & 3.9563 & 0.8094 \\
APCodec   \cite{ai2024apcodec}  & Spectral & 9 kbps    & 3.8749 & 3.1711 & 0.8937 & 0.9627 & 4.1762 & \underline{0.7945} \\
MDCTCodec \cite{jiang2024mdctcodec}  & Spectral & 9 kbps    & -      & -      & 0.9230 & -      & \underline{4.2600} & 0.8160 \\
STFTCodec (ours)  & Spectral & 9 kbps    & \underline{3.9716} & \textbf{4.1231} & \textbf{0.9648} & \textbf{0.9827} & \textbf{4.3405} & \textbf{0.7799} \\

\hline
DAC       \cite{kumar2024high}     & Waveform & 3 kbps    & \underline{3.8449} & \underline{2.9053} & \underline{0.8493} & \underline{0.9585} & 3.6002 & \underline{0.8495} \\
APCodec   \cite{ai2024apcodec}   & Spectral & 3 kbps    & 3.6100 & 2.5146 & 0.8354 & 0.9419 & \textbf{3.8756} & 0.8506 \\
STFTCodec (ours)  & Spectral & 3 kbps    & \textbf{3.8855} & \textbf{3.3756} & \textbf{0.8989} & \textbf{0.9714} & \underline{3.8302} & \textbf{0.8373} \\

\hline
DAC \cite{kumar2024high}     & Waveform & 1.5 kbps    & \underline{3.4595} & \underline{2.1401} & 0.7900 & \underline{0.9311} & 3.4916 & \underline{0.8722} \\
APCodec \cite{ai2024apcodec}   & Spectral & 1.5 kbps    & 3.3327 & 1.9432 & \underline{0.7982} & 0.9222 & \textbf{3.6799} & 0.8782 \\
STFTCodec (ours)  & Spectral & 1.5 kbps  & \textbf{3.6442} & \textbf{2.5457} & \textbf{0.8240} & \textbf{0.9544} & \underline{3.6653} & \textbf{0.8604} \\

\hline
\end{tabular}
\end{table*}

\subsection{Training Criteria}

\subsubsection{Mel-spectogram loss}
Following DAC \cite{kumar2024high}, we employ a multi-scale mel-spectrogram loss (\(\mathcal{L}_{mel}\)) using window lengths of 2048 and 512 samples, with hop sizes set to one-quarter of the window length. The loss combines both direct magnitude differences and log-scale differences using L1 distance.

\subsubsection{GAN-based loss}
We implement an adversarial training scheme based on the least-squares GAN framework. The discriminator produces multi-scale feature representations, and the generator training incorporates both adversarial feedback (\(\mathcal{L}_{adv}\)) and feature matching (\(\mathcal{L}_{feat}\)) to stabilize training and improve output quality.

\subsubsection{Codebook learning loss}
The codebook learning combines a vector quantization loss (\(\mathcal{L}_{vq}\)) minimizing the L1 distance between encoder output and quantized vectors, and a commitment loss (\(\mathcal{L}_{commit}\)) preventing excessive encoder output growth. Both losses employ stop-gradient operators to ensure proper gradient flow.

\subsubsection{Total Training Objective}
While our model leverages magnitude and phase spectra as intermediate representations, we do not explicitly enforce their reconstruction like APCodec \cite{ai2024apcodec}. Instead, we optimize for end-to-end perceptual quality through a weighted combination of the above losses:

\begin{equation}
    \mathcal{L}_{total} = \lambda_1\mathcal{L}_{mel} + \lambda_2\mathcal{L}_{feat} + \mathcal{L}_{adv} + \lambda_3\mathcal{L}_{commit} + \mathcal{L}_{vq}
\end{equation}

\noindent where \(\lambda_1=15.0\), \(\lambda_2=2.0\), and \(\lambda_3=0.25\). As demonstrated in \tabref{tab:ablation}, this design choice of avoiding explicit spectral reconstruction constraints allows the model to learn more perceptually optimal representations, resulting in superior audio quality.

\section{Experiments}
\label{sec:experiments}

\subsection{Datasets and Implementation Details}
We conducted experiments on two datasets: (1) VCTK Corpus \cite{veaux2017cstrvctk}, containing approximately 44 hours of high-quality speech data sampled at 48kHz, and (2) LibriTTS \cite{zen2019libritts}, a widely used speech dataset sampled at 24kHz. For VCTK, we randomly selected 400 utterances as the test set and use the remaining data for training. For LibriTTS, we used the standard test-clean subset for evaluation.

The input audio was transformed using STFT with a window size of 320 samples, hop size of 40 samples, and FFT size of 1024 points to obtain amplitude and phase spectra. The encoder architecture consisted of three downsampling layers, each with a downsampling factor of 2, resulting in a total downsampling ratio of 320 (40 $\times$ 8). This matched the compression ratio of existing neural codecs. To ensure fair comparison with baseline models like DAC \cite{kumar2024high} that use larger strides (512), we implemented a variant with a hop size of 80, extending the effective receptive field to 640 (80 $\times$ 8).

For the 12kbps configuration, we employed 8 codebooks for RVQ, each containing 1024 codewords of dimension 8. Lower bitrate variants were achieved by reducing the number of codebooks while maintaining other parameters. During training, audio inputs were segmented into approximately 0.33-second chunks (15,960 samples at 48kHz). All models were trained on a single NVIDIA RTX A6000 GPU with a batch size of 64. We optimized the model using AdamW with an initial learning rate of 5e-5 and momentum parameters \(\beta_1=0.8\) and \(\beta_2=0.99\). The learning rate was scheduled to decay exponentially with a decay rate of 0.999.

\begin{table}[!t]
\centering
\caption{Comparison on LibriTTS test-clean subset at 6 kbps.}
\label{tab:comparisonLibriTTS}
\setlength{\tabcolsep}{4.5pt}
\begin{tabularx}{\linewidth}{lccccc}
    \hline
    \textbf{Model} & \textbf{PESQ $\uparrow$} & \textbf{STOI $\uparrow$} & \textbf{VISQOL $\uparrow$} & \textbf{LSD $\downarrow$} & \textbf{MUSRHA $\uparrow$} \\
    \hline
    GT & - & - & - & - & 90.85$\pm$3.34\\
    APCodec    & 3.0601 & 0.9537  & \underline{4.3964} & \textbf{0.8327} & 74.88$\pm$4.21 \\
    DAC        & \underline{3.3297} & \underline{0.9549}  & 4.1239 & 0.8932 & \underline{82.49$\pm$3.49} \\
    STFTCodec  & \textbf{3.5289} & \textbf{0.9622}  & \textbf{4.4319} & \underline{0.8336} & \textbf{88.65$\pm$3.05} \\
    \hline
\end{tabularx}
% \vspace{-1.3em}
\end{table}

\subsection{Baselines and Evaluation Metrics}

We compared our method with several neural audio codecs: (1) AudioDec \cite{wu2023audiodec}, which integrates an encoder with HiFi-GAN vocoder for 48 kHz audio; (2) DAC \cite{kumar2024high}, the current SOTA model combining RVQ with low-dimensional VQ for waveform features; (3) APCodec \cite{ai2024apcodec}, processing amplitude and phase spectra with RVQ quantization; (4) MDCTCodec \cite{jiang2024mdctcodec}, using MDCT spectrum encoding with RVQ and multi-resolution discriminator.

We evaluated the codec performance using the following metrics: (1) UTMOS \cite{saeki2022utmos}, an automatic MOS prediction system that simulates human perceptual evaluation through machine learning models; (2) PESQ \cite{rix2001perceptual}, a standardized metric for speech quality assessment that quantifies distortion between processed and reference signals; (3) STOI \cite{taal2011algorithm}, evaluating speech intelligibility by analyzing short-time spectral features; (4) V/UV F1, measuring the classification accuracy of voiced and unvoiced speech segments; (5) ViSQOL \cite{chinen2020visqol}, a perceptual quality metric simulating human auditory perception; and (6) Log-spectral distance (LSD), calculating the distance between log-mel spectrograms of reconstructed and ground truth waveforms. For subjective evaluation, we conducted MUSHRA listening tests \cite{series2014method} with 15 trained listeners on the LibriTTS dataset \cite{zen2019libritts}, where perceptual differences between codecs are more distinguishable than VCTK \cite{veaux2017cstrvctk}. Participants evaluated 12 randomly selected test samples, and responses with reference scores below 80 are excluded to ensure reliability. We report both the mean scores and 95\% confidence intervals for the MUSHRA results in \tabref{tab:comparisonLibriTTS}.

\begin{table}[!t]
\centering
\caption{Comparison of STFTCodec variants against APCodec and DAC.}
\label{table:different_hop}
\setlength{\tabcolsep}{2pt}
\begin{tabularx}{\linewidth}{lrccccc}
    \hline
    \textbf{Model} & \textbf{Params} & \textbf{Stride $\uparrow$} & \textbf{PESQ $\uparrow$} & \textbf{STOI $\uparrow$} & \textbf{V/UV F1 $\uparrow$} & \textbf{ViSQOL $\uparrow$} \\
    \hline
    APCodec & 17.26M & 320 & 3.2155 & 0.9024 & 0.9642 & \underline{4.2240} \\
    DAC & 76.73M & \underline{512} & \underline{4.1464} & \underline{0.9565} & \underline{0.9804} & 4.1309 \\
    (H80, R8)$^{\mathrm{a}}$ & 36.87M & \textbf{640} & \textbf{4.2286} & \textbf{0.9732} & \textbf{0.9863} & \textbf{4.3142} \\
    \hline
    (H40, R8) & 41.98M & 320 & \textbf{4.0554} & \textbf{0.9670} & \textbf{0.9843}& 4.2305 \\
    (H80, R4) & 36.73M & 320 & 3.9800 & 0.9627 & 0.9817 & 4.1507 \\
    (H80, R8)$^{\mathrm{b}}$ & 36.87M & 640 & 4.0315 & 0.9658 & 0.9837 & \textbf{4.2996} \\
    \hline
    \multicolumn{7}{l}{$^{\mathrm{a}}$Model trained for approximately 200k steps, achieving better performance.} \\
    \multicolumn{7}{l}{$^{\mathrm{b}}$Early version of the same configuration trained for around 100k steps.} \\
    \multicolumn{7}{l}{Note: H and R denote STFT hop size and downsampling ratio respectively.}
\end{tabularx}
% \vspace{-1.3em}
\end{table}

\begin{table}[t]
\centering
\caption{Computational efficiency at 12kbps on VCTK dataset. Values in parentheses show speedup relative to real-time.}
\setlength{\tabcolsep}{5.2pt}
\begin{tabularx}{\linewidth}{lcccc}
    \hline
    \textbf{Model} & \textbf{Params} & \textbf{RTF (GPU) $\downarrow$} & \textbf{RTF (CPU) $\downarrow$} & \textbf{Training $\downarrow$} \\
    \hline
    DAC         & 76.73M & 0.064 (15.6$\times$) & 0.971 (1.02$\times$) & 1411.99 s/e \\
    APCodec     & 17.26M & \underline{0.010 (100$\times$)} & \textbf{0.115 (8.69$\times$)} & \underline{521.51 s/e} \\
    STFTCodec   & 36.87M & \textbf{0.007 (142$\times$)} & \underline{0.120 (8.33$\times$)} & \textbf{499.19 s/e}  \\
    \hline
\end{tabularx}
\label{tab:efficiency}
% \vspace{-1.3em}
\end{table}

\subsection{Main Results}
\tabref{table:codec_comparison} presents a comprehensive comparison between our proposed STFTCodec and other baseline models across different bitrates on the VCTK dataset \cite{veaux2017cstrvctk}. Our experiments demonstrate consistent advantages of STFTCodec in most evaluation metrics.

At 12 kbps, STFTCodec achieves the best performance in most metrics, particularly in perceptual quality (PESQ) and speech intelligibility (STOI). The performance advantage remains evident at 9 kbps, where STFTCodec outperforms other codecs in most metrics with only a marginal gap in UTMOS compared to DAC \cite{kumar2024high}. In ultra-low bitrate scenarios (3 kbps and 1.5 kbps), STFTCodec shows remarkable robustness and maintains its advantages across most metrics, demonstrating the effectiveness of our spectral-based approach under severe compression constraints.

To further validate the generalization capability, we conducted additional experiments on the LibriTTS dataset \cite{zen2019libritts} at 24kHz. As shown in \tabref{tab:comparisonLibriTTS}, STFTCodec achieves consistent advantages across all objective metrics at 6 kbps. Notably, our method demonstrates superior performance in both objective and subjective metrics, approaching the ground truth quality with a significantly smaller gap compared to baselines, while maintaining comparable spectral reconstruction quality to APCodec \cite{ai2024apcodec}. These results further validate STFTCodec's effectiveness in preserving speech quality.

These comprehensive results demonstrate that our method significantly outperforms existing spectral-based approaches while remaining competitive with state-of-the-art waveform-based methods across different bitrates and datasets, all while maintaining lower computational requirements.

\subsection{Flexibility Analysis}
An important advantage of our spectral-based approach is its ability to flexibly adjust the stride factor through STFT hop size modification, as opposed to waveform-based methods that require careful architectural designs of downsampling layers. \tabref{table:different_hop} presents the performance of different STFTCodec configurations with varying stride factors.

Our experiments show that STFTCodec maintains high performance across different configurations. Notably, the (H80, R8) variant achieves a stride factor of 640, significantly higher than DAC (512) and APCodec (320), while outperforming both models across all metrics. This demonstrates our method's unique capability to achieve higher compression ratios through simple parameter adjustment while preserving high audio quality.

\subsection{Computational Efficiency}
We conducted rigorous comparisons under controlled settings to evaluate computational efficiency. Due to the memory-intensive nature of DAC, training time was measured on RTX A6000 (48GB) with batch size 32 and 16 workers. Real-Time Factor (RTF) was benchmarked on AMD EPYC 7Y83 (CPU) and RTX 4090 (GPU) to demonstrate inference efficiency on consumer hardware. As shown in \tabref{tab:efficiency}, our method achieves superior efficiency while maintaining the best reconstruction quality.

\begin{table}[!t]
\centering
\caption{Ablation studies on key components of STFTCodec.}
\setlength{\tabcolsep}{3.4pt}
\begin{tabularx}{\linewidth}{lccccc}
    \hline
    \textbf{Model} & \textbf{PESQ $\uparrow$} & \textbf{STOI $\uparrow$} & \textbf{V/UV F1 $\uparrow$} & \textbf{ViSQOL $\uparrow$} & \textbf{LSD $\downarrow$} \\
    \hline
    STFTCodec      & \textbf{4.1436} & \textbf{0.9714} & \textbf{0.9852} & \textbf{4.3162} & \textbf{0.7796} \\
    w/o unwrap     & 4.1280 & 0.9684 & 0.9835 & 4.1912 & 0.7824 \\
    w/o ConvNeXt   & 3.8592 & 0.9486 & 0.9819 & 4.1948 & 0.8029 \\
    w/o MSTFTD     & 3.7254 & 0.9515 & 0.9832 & 3.9223 & 0.8216 \\
    w/ spectral recon. & 3.7102 & 0.9318 & 0.9778 & 4.1262 & 0.8073 \\
    \hline
\end{tabularx}
\label{tab:ablation}
\end{table}

\subsection{Ablation Studies}
We conducted ablation studies to validate key components of STFTCodec as shown in \tabref{tab:ablation}. We examined the impact of: (1) unwrapped phase derivatives, which enables faster convergence and better audio fidelity; (2) replacing ConvNeXt blocks with HiFi-GAN's well-established ResBlocks \cite{kong2020hifi} for feature extraction; (3) using only single-scale STFT discriminator and mel-spectrogram loss instead of multi-scale variants; and (4) adding explicit spectrum constraints similar to APCodec \cite{ai2024apcodec}. The results demonstrate that each component contributes to the model's performance, particularly validating our design choice of phase-aware reconstruction without strict spectral constraints.

\section{Conclusion}
We have presented STFTCodec, a spectral-based neural audio codec that effectively addresses the limitations of waveform-based approaches. Our key findings include: (1) spectral processing enables efficient compression without requiring large model capacity; (2) our parallel magnitude-phase processing architecture with unwrapped phase derivatives demonstrates superior perceptual quality; and (3) by relaxing strict spectral reconstruction constraints while maintaining phase-aware processing, we have achieved better audio quality than existing approaches. Experimental results validate our design and suggest promising directions for leveraging spectral transforms in neural audio coding and generation tasks.

% \newpage

\bibliographystyle{IEEEbib}
\bibliography{icme2025_fengtao}

\end{document}